# Superconductivity in Diamond-like $BC_3$ Phase: A First-principles Investigation

M.M. Ali, A.K.M.A. Islam[1], M. Aftabuzzaman, F. Parvin

*Department of Physics, Rajshahi University, Rajshahi-6205, Bangladesh*

**Abstract**

Two possible phases of superhard material $BC_3$ originating from the cubic diamond structure are investigated by *ab initio* pseudopotential density functional method using generalized gradient approximation (GGA). We calculate their elastic constants, electronic band structure, and density of states (DOS). Full phonon frequencies, electron-phonon coupling constant and possible superconducting $T_c$ of the metallic phase with tetragonal symmetry ($t$-$BC_3$, space group $P$-42m) have for the first time been investigated at 5 and 10 GPa. The calculated electron-phonon coupling (0.67) and the logarithmically-averaged frequency (862 cm$^{-1}$) show super-conductivity for the undoped $t$-$BC_3$ with $T_c$ = 20 K at 5 GPa, which decreases to 17.5 K at 10 GPa.

**Keywords:** Superhard $BC_3$, Band structure, Phonon spectra, Superconductivity

## 1. Introduction

Boron-rich diamond materials ($BC_x$) are of interest because these possess a unique combination of excellent physical and chemical properties such as very high hardness, low mass density, high mechanical strength, high thermal conductivity, excellent wear resistance, and high chemical inertness.[1] The synthetic and hypothetical superhard materials which belong to covalent and polar-covalent compounds are in growing importance in applications in mechanical machining, in electronic industry and solar cells production. The study of B as a hole dopant in diamond has a long history, there have been recent developments due to the ability to dope diamond films more heavily.[2] The strong B-B bonds in the graphitic layers of $MgB_2$ make it relevant in superconductivity, although graphite itself and diamond are materials that have even stronger bonds. Of the latter two, only diamond has bonding states that can conceivably become conducting through hole doping.[3] The question of mechanisms in strongly covalent materials is revived due to a recent report of superconductivity at 4 K in very heavily boron-doped diamond[4] and subsequent confirmation of $T_c \approx 7$ K in

[1]*E-mail address*: azi46@ru.ac.bd

B-doped diamond films.[5]

The interest in these types of compounds is continuing since the successful synthesis of a graphite-like layered structure of bulk $BC_3$ quite some time earlier[6] by the chemical reaction of benzene and boron trichloride at 800°C. The material is similar to 25% boron-doped diamond without boron clusters. The synthesized material was suggested to be of $BC_3$ layers with hexagonal symmetry.[7-9] But it was not clear about the way the layers are stacked in bulk material. The *ab initio* local-density approximation calculations[8,9] predicted that bulk $BC_3$ is metallic due to the interaction between the adjacent $BC_3$ layers. Ribeiro and Cohen[10] studied possible superconductivity in hole-doped $BC_3$ considering it as a layered material with ABC-stacking configuration with eight atoms per unit cell. It was shown that in such hypothetical structure superconductivity with $T_c \sim 22$ K could be achieved with doping levels of 02-0.3 holes/cell. Sun *et al.*[11] predicted two stable $BC_3$ structures, one semiconductor and the other metal using *ab initio* LDA pseudopotential density functional method. More recently Liu *et al.*[12] predicted a sandwich-like conducting tetragonal $BC_3$. The structure is predicted to be formed by alternately stacking sequence of metallic CBC block and insulating CCC block. This sandwich-like metal and/or insulator layered structure is shown to exhibit anisotropic conductivity on the basal planes formed by the metallic CBC blocks along the *c*-axis. The study also predicts synthesis of *t*-$BC_3$ from the graphite-like $BC_3$ (*g*-$BC_3$) at a critical pressure of only ~ 4 GPa.

The purpose of this work is to examine further the metallic behaviour *t*-$BC_3$ in comparison to *g*-$BC_3$ and possible superconductivity in the metallic phase. The superhard material is expected to have large elastic constants, requiring short and strong chemical bonds which are typically seen in light-element materials.[13] These would favour energetic phonon modes. Besides the elastic constants we would also compute full phonon frequencies and electron-phonon coupling which would allow us to estimate the superconducting transition temperature.

## 2. Computational methods

Our calculations were performed using the *ab initio* plane-wave pseudopotential approach within the framework of the density-functional theory implemented in the CASTEP software.[14] The ultrasoft pseudopotentials were used in the calculations, and the plane-wave cutoff energy was 310 eV for *t*-$BC_3$. The exchange-correlation terms were considered by the Perdew-Berke-Ernzerhof form of the generalized gradient approximation.[15] The *k*-points samplings were 7×7×7 and 4×8×6 in the Brillouin zone for the *t*-$BC_3$ and the graphitelike $BC_3$ (*g*-$BC_3$), respectively, according to the Monkhorst-Pack scheme. All the structures were relaxed by the Broyden-Fletcher-Goldfarb-Shanno methods.[16] The elastic constants $C_{ij}$, bulk modulus $B$ and electronic properties were directly calculated by the CASTEP code. The phonon calculation was carried out using the linear-response method with $q$ vector of 2×2×2 available in the Quantum ESPRESSO code.[17]

## 3. Results and discussion

### 3.1 Elastic properties

The structural minimization of the two predicted lowest energy configurations $t$-BC$_3$ (tetragonal, $P\bar{4}2m$, No. 111) and $g$-BC$_3$ (orthorhombic, $Cmcm$, No. 63) has been performed within the framework of the density-functional theory in order to utilize the results in our present analysis. The two structures are shown in Fig. 1 (a, b). There are 8 atoms and 32 atoms in the unit cell of the two structures, respectively. The equilibrium structural parameters, $a = b = 3.548$ Å, $c = 3.911$ Å and total energy -135.3254 eV/atom calculated for the t-BC$_3$ at ambient pressure are found to be in reasonable agreement with the earlier calculations.[12] On the other hand two ($a, c$) of the three lattice parameters for $g$-BC$_3$ structure at ambient pressure are found to be 7-13% higher in the present case compared to those obtained earlier.[12] It is further observed that the $g$-BC$_3$ phase keeps its layered structure even at pressure as high as 100 GPa. Fig. 1(c) shows the enthalpy difference of $t$-BC$_3$ with respect to $g$-BC$_3$ as a function of pressure. It is seen that at ambient pressure the energy difference between the phases is small and $t$-BC$_3$ is energetically favourable for $P \geq 5$ GPa.

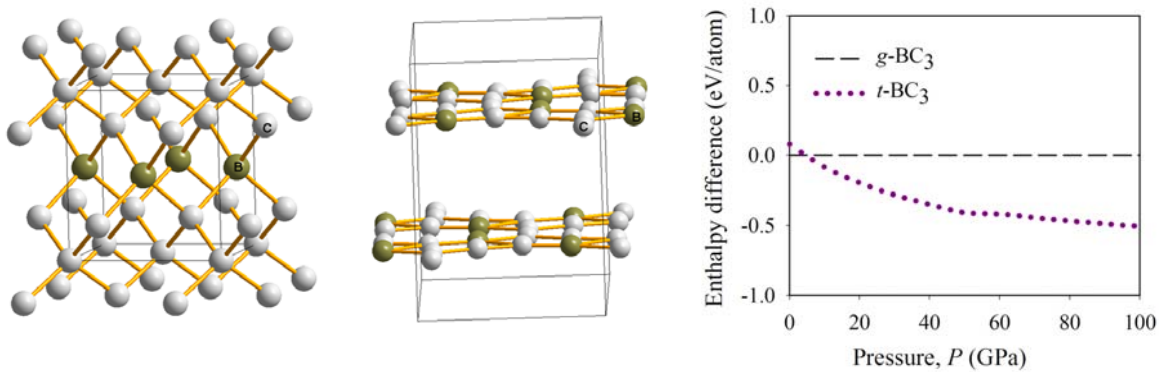

Fig. 1. Structure of (a) $t$-BC$_3$ (*left*), (b) $g$-BC$_3$ (*middle*) at ambient pressure and (c) Enthalpy difference between these phases as a function of pressure.

We have calculated the six independent elastic constants obtained by the finite strain techniques for $t$-BC$_3$. The variation of elastic constants as a function of pressure is shown in Fig. 2. As can be seen the values of all the elastic constants increase with the increase of pressure, except $C_{66}$ which despite some nonlinear variation remains nearly flat. These results do not show any evidence of drastic change in the behaviour with pressure. Our calculated elastic constants of $t$-BC$_3$ at $P = 0$ are all positive and satisfy the mechanical stability criteria for a tetragonal crystal.[18,19] Higher pressure-values are also seen to meet the above conditions.

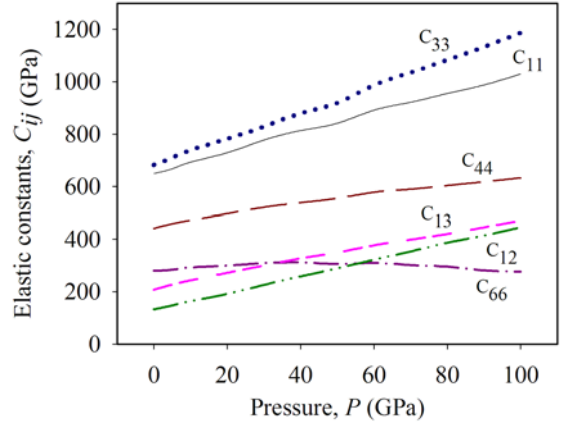

Fig. 2. Variation of elastic constants of $t$-BC$_3$ as a function of pressure.

The calculated nine elastic stiffness constants (in GPa) for $g$-BC$_3$ at $P = 10$ GPa are as follows: $C_{11} = 987$, $C_{22} = 985$, $C_{33} = 87$, $C_{44} = 21$, $C_{55} = 11$, $C_{66} = 390$, $C_{12} = 187$, $C_{13} = 12$, $C_{23} = 23$. All these satisfy the stability criteria for an orthorhombic crystal.[20], which indicate that $g$-BC$_3$ phase is also mechanically stable.

The theoretical polycrystalline elastic moduli for $t$-BC$_3$ may be calculated from the set of independent elastic constants. Hill[21] proved that the Voigt and Reuss equations represent upper and lower limits of the true polycrystalline constants. He showed that the polycrystalline moduli are the arithmetic mean values of the moduli in the Voigt ($B_R$, $G_R$) and Reuss ($B_R$, $V_R$) approximation, and are thus given by $B_H \equiv B = \frac{1}{2}(B_R + B_V)$ and $G_H \equiv G = \frac{1}{2}(G_R + G_V)$. The expression for Reuss and Voigt moduli can be found elsewhere.[22] The Young's modulus $Y$ and Poisson's ratio $\nu$ are then computed from these values using the relationship: $Y = 9BG/(3B+G)$, $\nu = (3B-Y)/6B$. As compared with other superconductors, $t$-BC$_3$ is a hard material. This is evident from the value of high bulk moduli 340-360 GPa for $t$-BC$_3$ at $P = 0$, which can be compared with the bulk moduli of 442 GPa for diamond[23] and 409 GPa for $c$-BN.[24] The shear modulus of $t$-BC$_3$ is 315 GPa in comparison to 336.5 GPa due to Liu $et\ al$.[12] It is known that the values of the Poisson ratio ($\nu$), minimal for covalent materials ($\nu = 0.1$), increase for ionic systems.[25)] In our case, the value of $v$ for $t$-BC$_3$ is 0.147, which is indicative of covalent contribution in intra-atomic bonding.

*3.2 Electronic properties*

The electronic band structures of the $t$-BC$_3$ phase at several pressures are shown in Fig. 3. It is observed that the top of the valence band is ~ 2.28 eV (~ 2.7 eV[12)]) above Fermi level at equilibrium pressure. The corresponding values at 5, 40 and 80 GPa are 2.31, 2.51 and 2.71 eV, respectively. That is, the valence band is shifted more with the increase of pressure. The result, for example at zero pressure, is at variance with covalent isoelectronic compounds (with the zinc-blend structure) that are semiconducting, i.e., the tops of their valence bands are tangential to the Fermi level.

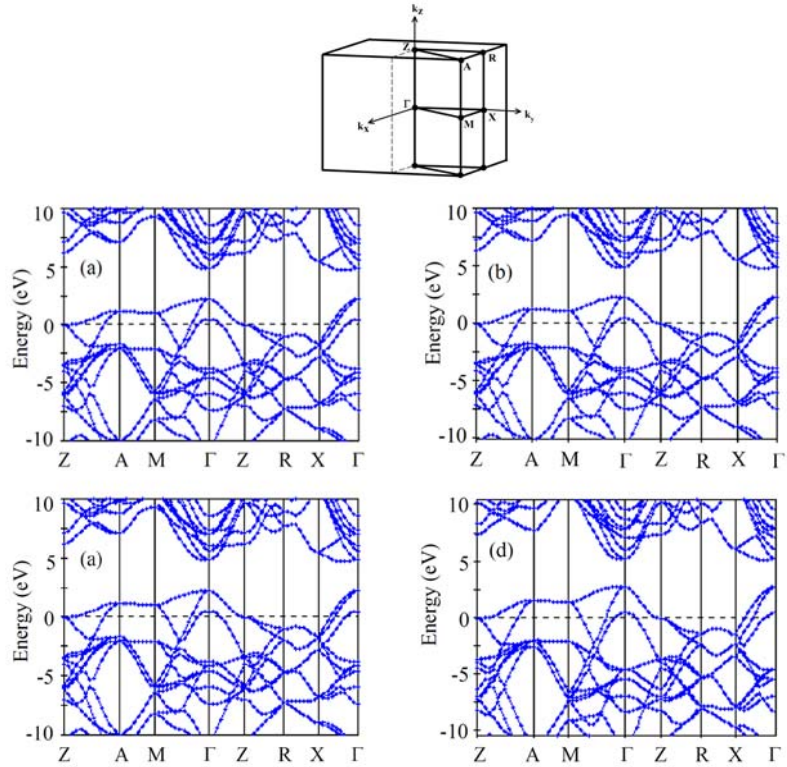

Fig. 3. Band structures of $t$-BC$_3$ at (a) $P$=0, (b) $P$= 5, (c) $P$= 40, (d) $P$= 80 GPa. (*Top*) First Brillouin zone.

On looking at the band structure for $t$-BC$_3$ further we see that some empty orbitals appear above the Fermi level. This is simply because there is electron deficiency of B atoms in the system. The charges are not balanced and the electrons do not fill up the $sp^3$-hybridized orbitals. Further there is an absence of a gap at the Fermi level also.

The appearance of empty bands above Fermi level in the *t*-BC$_3$ (see Fig. 3) is thus quite different from the semiconducting isoelectronic compounds. The existence of these incompletely filled valence bands implies that *t*-BC$_3$ exhibits metallicity. Liu *et al.*[12] found similar band structure in their study. Further in an earlier report He *at al*[26] found similar characteristics in the band structure of the electron-deficient B$_2$CN, which has been predicted to have the conducting property.

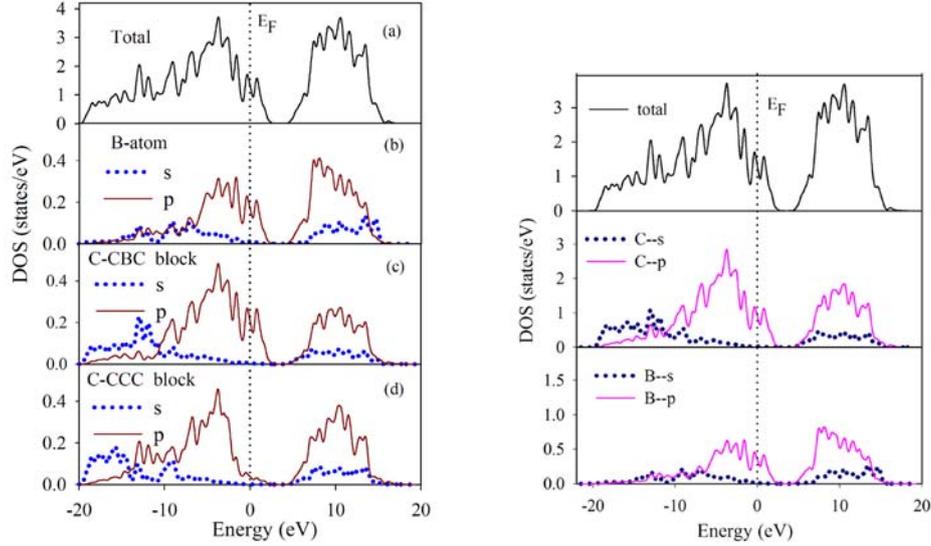

Fig. 4. (*Left panel*): (a) The total density of states for *t*-BC$_3$ and local density of states for the (b) B atom, (c) C atom in the CBC block, and (d) C atom in the CCC block at equilibrium pressure.(*Right panel*): The total and partial DOS for *t*-BC$_3$.

In Fig. 4 we show the calculated total and partial density of states for *t*-BC$_3$. In the left panel of the figure are shown (a) the total density of states, and the local density of states for the (b) B atom, (c) C atom in the CBC block, and (d) C atom in the CCC block at equilibrium pressure. The partial DOS is shown in the bottom two figures on the right panel. The contribution of *s* and *p* local density of states for each of C and B atoms in the *t*-BC$_3$ unit cell is clearly seen in the figure. It is observed that the valence bands beneath the Fermi level are mainly due to the 2*p* orbitals of the C and B atoms. On the other hand the empty bands above the Fermi level originate almost equally from the 2*p* orbitals of B and C atoms in the CBC block with a small contribution from 2*p* orbitals of C atoms in the CCC block. The empty orbitals are persistent to the $sp^3$-hybridized B-C bonds in the CBC block. This would indicate that the CBC block is metallic, while the CCC block is almost insulating. As a result, the conduction in the *t*-BC$_3$ crystal would proceed anisotropically on its basal planes formed by the CBC, rather than the CCC blocks.

The band structures of $g$-BC$_3$ crystal at $P = 0$ and 10 GPa are shown in Fig. 5. As pressure increases from zero to 10 GPa the equilibrium lattice parameters $a$ and $c$ change by 7% and 13%, respectively, whereas $b$ changes slightly. As a result the Γ-Z width (c-direction), for example, increases in the reciprocal space as can be observed in Fig. 5. We observe that there are no empty bands above the Fermi level and the tops of the valence bands are tangential to the Fermi level. The band structures look to be of similar character for covalent isoelectronic compounds (with the zinc-blend structure) that are semiconducting. The features of the states near the Fermi level thus do not show any metallicity in $g$-BC$_3$.

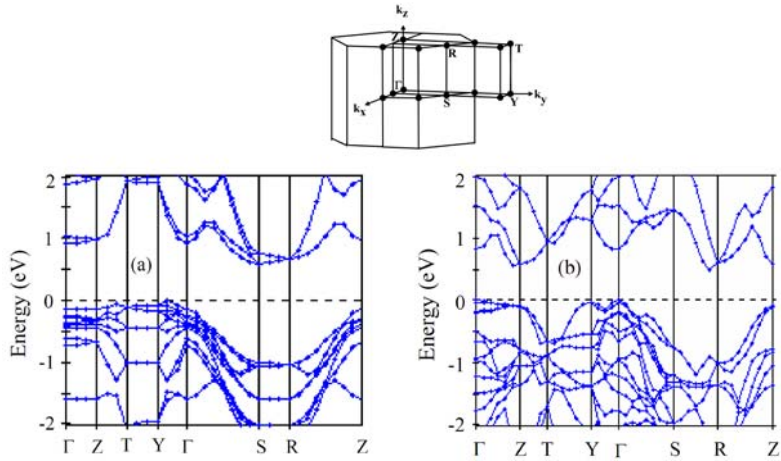

Fig. 5. Band structure of $g$-BC$_3$ at (a) $P = 0$ and (b) $P = 10$ GPa. (*Top panel*) First Brillouin zone.

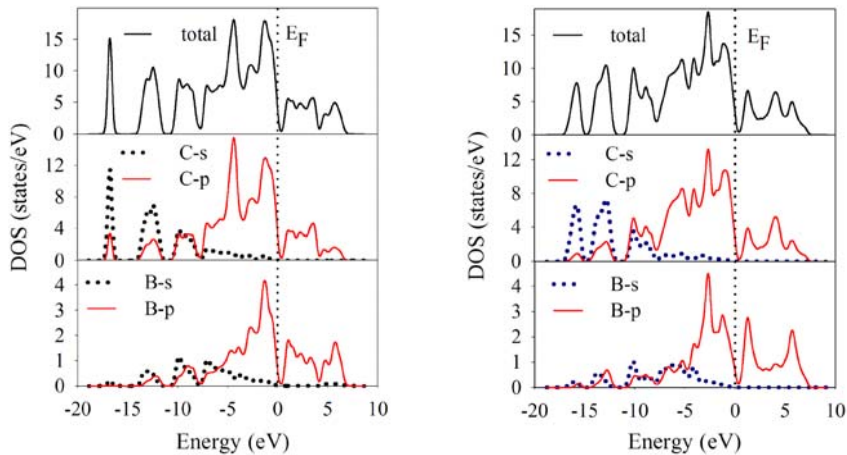

Fig. 6. Density of states of $g$-BC$_3$ at (a) $P = 0$ and (b) $P = 10$ GPa.

As shown in Fig. 6, the $s$ and $p$ local DOS for each atom (C, B) as well as the total

DOS in the *g*-BC$_3$ unit cell with 32 atoms with the *Cmcm* symmetry have been calculated. From this the contributions to DOS of *p* state to both C and B atoms are evident.

*3.3 Phonon spectrum and electron-phonon coupling*

We have calculated phonon dispersions of the metallic *t*-BC$_3$ at a pressure of 5 and 10 GPa. Fig. 7 shows only the results at 10 GPa. The corresponding phonon density of states is shown on the right panel of the figure. It is clear that all the phonon frequencies are real and no imaginary phonon frequency is observed in the whole Brillouin zone in the calculated phonon dispersion curve and phonon density of states of *t*-BC$_3$. This is also true for data at 5 GPa. This indicates that *t*-BC$_3$ structure is dynamically stable at both pressures. The phonon spectrum for 10 GPa extends up to 1110 cm$^{-1}$; it shows three main peaks centred at ~ 1060 cm$^{-1}$, ~ 990 cm$^{-1}$, and ~ 670 cm$^{-1}$. There are a few other smaller peaks also. The phonon vibrations when decomposed into atomic components show that, despite B being lighter than C, the harder phonon modes (990 - 1060 cm$^{-1}$) are due to C vibrations. The same features are observed by Calandra and Mauri[13] in diamond-like BC$_5$. The much weaker signal in the low frequency part (lower than 500 cm$^{-1}$) indicates very weak electron acoustic phonon coupling. In the optical range lying above this frequency but below the harder phonon modes, B related phonon vibrations are dominant. The B related phonon modes play important role in the e-ph coupling due to the large B electronic local DOS near the Fermi level.

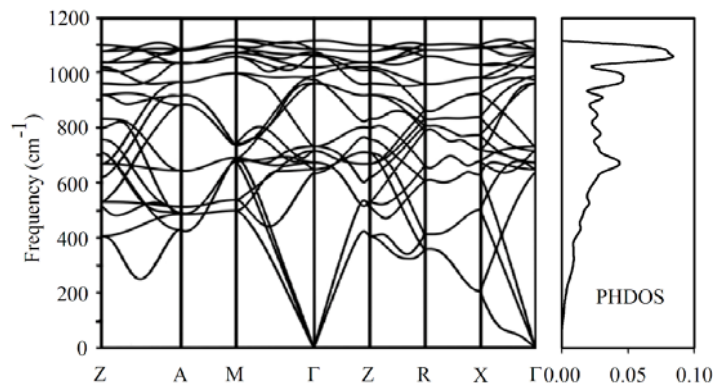

Fig. 7. Phonon dispersion curve (*left*) and phonon DOS (*right*) of *t*-BC$_3$ at 10 GPa.

The frequency-dependent electron-phonon coupling constant $\lambda(\omega)$ and Eliashberg function $\alpha^2 F(\omega)$ are given by[27]

$$\lambda(\omega) = 2 \int_0^\omega d\Omega\, \alpha^2 F(\Omega)/\Omega \tag{1}$$

$$\alpha^2 F(\omega) = \frac{1}{N(0)} \sum_{nmk} \delta(\varepsilon_{nk}) \delta(\varepsilon_{mk+q}) \sum_{vq} |g_{v,nk,m(k+q)}|^2 \delta(\omega - \omega_{vq}) \tag{2}$$

where $g_{v,nk,mk'}$ are the bare DFT matrix elements.

The total e-ph coupling constant $\lambda$ for metallic state of $t$-BC$_3$, is obtained by numerical integration of equation (1) up to $\omega = \infty$, which yields $\lambda = 0.67$ (5 GPa) and 0.66 (10 GPa). The corresponding values for the logarithmically-averaged frequencies ($\omega_{ln}$) are calculated to be 862 and 763 cm$^{-1}$, respectively. We use the Coulomb pseudopotential $\mu^* = 0.13$ and the simplified Allen-Dynes formula for $T_c$[28] within the phonon mediated theory of superconductivity:

$$k_B T_c = \frac{\hbar \omega_{ln}}{1.2} \exp\left[ -\frac{1.04(1+\lambda)}{\lambda - \mu^*(1+0.62\lambda)} \right] \tag{3}$$

This yields $T_c = 20$ K (5 GPa) and 17.5 K (10 GPa) for the undoped $t$-BC$_3$ formed by alternately stacking sequence of metallic CBC block and insulating CCC block. This shows that undoped $t$-BC$_3$ with such a stacking is a phonon-mediated superconductor. Ribeiro and Cohen[10] have shown that for metallic $t$-BC$_3$ structure with ABC-stacking, superconductivity at $T_c \sim 22$ K could be achieved with doping levels of 0.2-0.3 holes/cell. However they pointed out that their $T_c$ may have been overestimated due to the approximate method followed and in case electron-phonon coupling is peaked at $\Gamma$. This is because they did not calculate phonons for the entire Brillouin zone, rather they calculated only the zone-centre phonons. It is worth mentioning that the calculated DOS at the Fermi level decreases with pressure (15% decrease over a pressure range of 100 GPa), which would imply that $T_c$ of $t$-BC$_3$ would decrease with pressure. This is in agreement with our calculated $T_c$-values mentioned above.

## 4. Conclusions

The structural stability of superhard $t$-BC$_3$ with space group $P\bar{4}2m$ (constructed by an alternately stacking sequence of metallic CBC and insulating CCC blocks along the $c$ axis) has been confirmed through our calculations of mechanical properties and full phonon frequencies. The $t$-BC$_3$ crystal in comparison to $g$-BC$_3$ shows metallicity at ambient and higher pressures because of the features of the states near the Fermi level.

The sandwich-like metal insulator lattice of $t$-BC$_3$ with relatively high symmetry but low isotropy is suggestive of conducting property on the basal planes formed by the CBC blocks. Our present results indicate for the first time that the undoped $t$-BC$_3$ is a phonon mediated superconductor with $T_c$ of 20 K and 17.5 K at 5 and 10 GPa, respectively.

The synthesis of $t$-BC$_3$ from $g$-BC$_3$ at a pressure above 5 GPa (range of energetically favourable $t$-phase) would seem to be difficult in view of the calculated mechanical stability and retention of the layered structure of $g$-BC$_3$ at higher pressure. The energy difference ($\Delta E$) between the semiconductor structure $g$-BC$_3$ and the metallic structure $t$-BC$_3$ is positive and small up to ~ 4 GPa, but above 5 GPa the difference becomes negative which increases continuously with pressure. Thus a proper catalyst could possibly initiate such a synthesis.

## Acknowledgment

The authors acknowledge the help provided by Rajshahi University.